\documentstyle[preprint,aps,graphicx]{revtex}
\tightenlines
\begin{document}

\title{Electronic and Transport Properties of Carbon Nano Peapods}

\author{A. Rochefort$^{*,\dagger}$}

\address{$^*$ \'Ecole Polytechnique de Montr\'eal, D\'epartement de g\'enie physique,\\
Montr\'eal, (Qu\'e) Canada H3C 3A7.} 
\address{$^\dagger$ CERCA, Groupe Nanostructures, Montr\'eal, (Qu\'e) Canada H3X 2H9.}

\maketitle

{\centerline{(\today)}}

\begin{abstract}
We theoretically studied the electronic and electrical properties of metallic and semiconducting peapods with encapsulated C$_{60}$ (C$_{60}$@CNT) as a function of the carbon nanotube (CNT) diameter. For exothermic peapods (CNT diameter $>$ 11.8 {~\AA}), only minor changes, ascribed to a small structural deformation of the nanotube walls, were observed. These include a small electron charge transfer (less than 0.10 electron) from the CNT to the C$_{60}$ molecules and a poor mixing of the  C$_{60}$ orbitals with those of the CNT. Decreasing the diameter of the nanotube leads to a modest increase of the charge density located between the C$_{60}$'s. More significant changes  are obtained for endothermic peapods (CNT diameter $<$ 11.8 {~\AA}). We observe a large electron charge transfer from C$_{60}$ to the tube, and a drastic change in electron transport characteristics and electronic structure. These results are discussed in terms of $\pi$-$\pi$ interaction and C$_{60}$ symmetry breaking.
\end{abstract}


\section{introduction}

The possibility of using new forms of carbon-based materials for practical applications in nanoelectronics has stimulated an important amount of exciting works during the last decade. \cite{dressel-avouris} Since their discovery in 1991, carbon nanotubes (CNTs) have been used as an active component in the fabrication of transistor \cite{tans,martel,postma}, memory elements \cite{rueckes}, and more recently, logic circuits.\cite{derycke,bachtold} CNT's can also be used as a template for nanofabrication and as reservoirs for the storage of gas, ions, or metals.\cite{1d-metals} In this respect, it has been recently shown experimentally that multiple C$_{60}$ molecules can penetrate into a carbon nanotube to form a one-dimensional array of C$_{60}$ nested inside.\cite{smith} This new type of carbon materials, due to its original structure, is often called carbon peapod.\\

Filling CNTs with C$_{60}$ is exothermic or endothermic depending on the size of the nanotube.\cite{okada} C$_{60}$@(10,10) was found to be stable (exothermic) while other peapods with a smaller CNT shell such as the (9,9) and (8,8) tubes are endothermic. Metallic CNTs preserve most of their intrinsic properties upon the encapsulation of C$_{60}$ molecules. The interaction between C$_{60}$ and the nanotube occurs through a weak orbital mixing between a near free electron (NFE) state on the CNT located above Fermi level (and most probably above vacuum level) and the $p$ orbitals of the C$_{60}$. This interaction leads to a weak electron confinement between C$_{60}$ and the CNT wall, and a slight charge transfer from the tube to the C$_{60}$.  Therefore, only little perturbation is expected. However, recent STM results show drastic modification of the local electronic structure of semiconducting nanotube peapods.\cite{yazdani} This perturbation is essentially present in the conduction band. The main variation observed is a sharp increase in the density of states probed directly over an encapsulated C$_{60}$ molecule. This increase in the DOS was tentatively attributed to the electronic coupling of C$_{60}$ with the CNT shell, which is estimated approximately to about 1 eV.\cite{yazdani} This tube-C$_{60}$ coupling appears much stronger than anticipated by theory \cite{okada} or experimental work on metallic peapods.\cite{liu}  It is therefore important to study further the properties of encapsulated C$_{60}$ on the electrical and electronic properties of both metallic and semiconducting nanotubes as a function of the nanotube diameter. In the present study, we show that the electronic as well as the electron transport properties of exothermic peapods do not show drastic differences from the properties of individual species. The CNT-C$_{60}$ interaction increases slightly as the nanotube diameter decreases, but becomes very important for endothermic peapods.

\section{computational details}

We considered the encapsulation of up to three C$_{60}$ molecules in approximately 10~nm long metallic and semiconductor CNT models containing up to 1500 carbon atoms. The C$_{60}$ molecules were systematically placed in the middle sections of the CNT. The region where C$_{60}$'s are encapsulated is relatively small with respect to the entire CNT length, thus avoiding the influence of the open boundary conditions implicit to our finite model. The computed stabilization energy ($\Delta E_S$) corresponds to the energy difference between a fully optimized peapod structure and the isolated species (perfect CNT and isolated C$_{60}$'s). Prior to the optimization, the dangling bonds at both ends of the finite nanotube models were saturated with hydrogen. The activation energy needed to introduce the C$_{60}$'s in the CNT was not evaluated. Stabilization energies were calculated with the MM3 molecular mechanic force field \cite{mm3} (the bond parameter of alkene was modified to 1.42 {\AA}) and the geometries were optimized with a standard conjugated gradient technique down to a root mean square (RMS) deviation $<$ 10$^{-5}$. We previously showed that this modified-MM3 force field gives a total energy that is in good agreement with the more accurate TB-DFT method \cite{rochefort_prb}. The electronic structure calculations for the CNT and peapod systems were carried out within the extended H\"uckel (EH) method, which includes an explicit treatment of overlap integral for the \emph{s} and \emph{p} valence orbital of carbon.\cite{yaehmop} It has been shown that EH gives results similar to those obtained on extended CNTs with more sophisticated  methods.\cite{rochefort_jpc}\\

The electrical transport properties of carbon nanotubes and peapods were computed using a Green's function approach \cite{datta} within the Landauer-B\"uttiker formalism. The Hamiltonian and overlap matrices used in this formalism were also determined using the EH model.\cite{yaehmop} For transport calculations,  the two ends of the CNT were bonded to gold electrodes. The metallic contacts consist of a sufficient number of gold atoms in a (111) crystalline arrangement to create a larger contact area relative to the CNT ends.  In order to minimize the contact resistance, the distance between the gold pad and the tube end was fixed to 1.5~\AA.\cite{rochefort_prb} The Green's function of the conductor can be written in the form of block matrices separating explicitly the molecular Hamiltonian:

\begin{equation}
G_C =\big[ E{\mathcal{S}}_C - {\mathcal{H}}_C -\Sigma_1-\Sigma_2 \big]^{-1}
\end{equation}

\noindent
where ${\mathcal{S}}_C$ and ${\mathcal{H}}_C$ are the overlap and the Hamiltonian matrices of the conductor (nanotube or peapod), respectively, and $\Sigma_{1,2}$ are self-energy terms describing the effect of the leads. The transmission function $\bar{T}(E)$ (or transmittance), which is the summation of transmission probabilities over all conduction channels in the system, is obtained from the Green's function of the conductor ($G_C$) given by
\cite{datta}:

\begin{equation} 
\bar{T}(E)=\bar{T}_{21}=\textrm{Tr} [\Gamma_2 ~G_C ~\Gamma_1~ G_C^\dagger ]
\end{equation} 

In this formula, the matrices have the form:  

\begin{equation} 
\Gamma_{1,2}=i(\Sigma_{1,2}-\Sigma^\dagger_{1,2})
\end{equation} 

\section{results and discussion}

The existence of encapsulated C$_{60}$ inside a CNT can be discussed in term of stabilization energy ($\Delta E_S$) in the form of:

\begin{displaymath}
\Delta E_S \equiv E\big[x \textrm{C}_{60}@(n,m)\big] - E\big[(n,m)\big] - x E\big[C_{60}\big]  \end{displaymath}

\vskip0.3cm
\noindent
where $x$ is the number of C$_{60}$ molecules, and $(n,m)$ is the chirality index of the nanotube considered. The variation of the stabilization energy as a function of the nanotube diameter (D$_{NT}$) for a single encapsulated C$_{60}$ molecule is shown in Figure 1, while Table 1 gives $\Delta E_S$ (reported by C$_{60}$) values for multiple encapsulated C$_{60}$ molecules. In order to qualitatively describe the influence of the $\pi$-electron cloud on the resulting peapods stability, we show an additional curve (the dotted line) in Figure 1 where the thickness of the $\pi$-cloud ($\approx$ 3.3~{\AA}) is substracted from the nanotube diameters. The stabilization energy for peapods with a large diameter CNT shell is relatively weak and exothermic. On the other hand, C$_{60}$ can more easily penetrate into large CNT because the interaction energy is weak. As the CNT diameter decreases, the interaction energy between the C$_{60}$ molecule and the CNT wall becomes more important; the $\Delta E_S$ values reflect then the balance between van der Waals attraction and Coulomb repulsion. The most stable peapod is the $x$C$_{60}@(10,10)$ system, in which the $\pi$-electron clouds of both C$_{60}$ and CNT just begin to overlap. For CNT smaller than the (10,10) tube, the peapods become rapidly less stable and highly endothermic (i.e. positive $\Delta E_S$) at D$_{NT} <$ 11.9~{\AA}. The range of tube diameters where peapods are the most stable (12 $<$ D$_{NT} <$ 15~{\AA}) is in good agreement with recent experimental \cite{bandow} and theoretical \cite{okada,qian} observations. The C$_{60}$@(9,9) peapod was previously found \cite{okada} slightly endothermic (by 6 kcal/mol) with DFT-LSD method. The lower stability found for C$_{60}$@(9,9) with DFT is probably related to the imposed commensurate structure of the C$_{60}$ within the tube in their supercell model. The relative stability of small peapods is strongly related to the ability of the tube to satisfy the presence of encapsulated C$_{60}$ molecules through a deformation of its structure, especially near the C$_{60}$ molecules. As expected, the smaller is the CNT, the larger are the structural deformations, and consequently lower is the stability of the peapod.\cite{okada,qian} This also reflects on the C$_{60}$-C$_{60}$ distance ($d_{C_{60}-C_{60}}$); C$_{60}$ molecules in small CNTs become squashed and the distance between their centers increases (see Table 1). In the following, we first compare the electronic and electrical properties of the most stable metallic (C$_{60}@(10,10)$) and semiconducting (C$_{60}@(16,0)$) peapods. Then, we present the results for the highly endothermic C$_{60}@(14,0)$ peapod.\\

Figure 2 compares the density of states (DOS) of perfect-CNT/peapods systems (upper panels), and the local density of states (LDOS) of the CNT shell of these peapods (lower panels) in a region near the C$_{60}$, for the (A) metallic 3C$_{60}$@(10,10) and (B) semiconducting 3C$_{60}@(16,0)$ structures. The vertical dotted-dashed lines indicate the energy position where a difference in LDOS was observed between a perfect CNT and a peapod for which the contribution of the C$_{60}$ molecules was removed. This last comparison allows us to highlight the changes induced by a structural deformation rather than by an electronic influence of C$_{60}$ on the CNT. The finite DOS at the Fermi energy ($E_F$) for C$_{60}$@(10,10), and the absence of states at $E_F$ for C$_{60}$@(16,0) peapod suggest that the fundamental (metal, semiconductor) electronic characteristics of the CNTs are preserved upon C$_{60}$ encapsulation. The presence of C$_{60}$ in the peapods clearly results in additional peaks in DOS for valence and conduction bands of both metallic and semiconducting peapods. The first two peaks near $E_F$ associated to C$_{60}$ are observed at around -1.0 eV and +0.2 eV for 3C$_{60}$@(10,10), and at around -0.7 eV and +0.5 eV for 3C$_{60}$@(16,0). These peaks correspond respectively to the HOMO and the LUMO of C$_{60}$. The calculated HOMO-LUMO gap (1.2 eV) for encapsulated C$_{60}$ is smaller than the value calculated for isolated C$_{60}$ (1.6 eV), and the experimental values ($\approx$ 1.6-1.8 eV) for C$_{60}$ in gas and solid phases.\cite{c60_gap} This difference is partly related to the deformation of C$_{60}$ which contributes to closing the gap of C$_{60}$,\cite{joachim} and to the displacement of C$_{60}$ orbitals induced by the relatively weak interaction between C$_{60}$ and the CNT. In addition, the very small charge transfer \cite{note} from the CNT to the C$_{60}$ molecule (see Table 1), and the weak mixing of states between C$_{60}$ and the CNT (C$_{60}$ states are weakly spread in the DOS for peapods) support that only weak interaction between C$_{60}$ and the nanotube wall are present. This result is in agreement with previous DFT-LSD description of metallic peapods, in which a very weak charge transfer was observed.\cite{okada} The influence of the CNT diameter on the charge density distribution is represented in Figure 3 in which we consider the residual charge density $\rho_r$ such as:

\begin{displaymath}
\rho_r = \rho[x\textrm{C}_{60}@(n,m)] - \rho[x\textrm{C}_{60}] - \rho[(n,m)] \end{displaymath}

\vskip0.3cm
\noindent
and where the dark and bright regions indicate a gain and a loss of electron charges density, respectively. As the CNT diameter decreases, the accumulation of charge density between the C$_{60}$ molecules, and between the C$_{60}$ and the CNT shell increases, leaving the C$_{60}$ molecules slightly more negative than the isolated species. This result contrasts slightly with previous DFT result where the weak accumulation of negative charge density in large diameter peapods was mainly located in the space between the tube and C$_{60}$.\cite{okada} This charge density localization, which is practically absent for peapods smaller than the (10,10) tube, was attributed to the presence of a weak coupling between C$_{60}$ and a near free electron (NFE) state of the CNT, which is known to be poorly described within EH method. The case of C$_{60}$@(14,0) peapod (D) is quite different. There is an important loss of charge density between the C$_{60}$ and the CNT shell and a small gain of charge density between C$_{60}$'s. As discussed below, this behavior is mainly related to C$_{60}$ symmetry breaking.\\

The LDOS of peapods (lower panels of Figure 2) also suggests a weak influence of the C$_{60}$ molecules on the electronic properties of the CNT peapods. In these LDOS diagrams, the label ``0" marks the carbons that are the closest to C$_{60}$, and the labels $\pm$1, $\pm$2, $\ldots$ are for carbon sections progressively away from the central ``0" position. For the metallic 3C$_{60}$@(10,10) peapod (see Figure 2A), the most important changes occur in the valence band between -1.7 and -1.1 eV. These variations are more directly related to the states created by structural deformation of the CNT shell (indicated by the vertical dashed-dotted lines) as opposed to the possible electronic influence of C$_{60}$. A similar situation occurs for the semiconducting 3C$_{60}$@(16,0) peapod, except that a smaller tube diameter results in slightly higher electronic influence on the LDOS of the (16,0) tube. A certain number of structural deformed states (dashed-dotted lines) coincides with the presence of C$_{60}$ states and suggest a possible mixing of states. In contrast to the DOS where the contribution of the tube and the three C$_{60}$ molecules are convoluted, the local variation of DOS in the vicinity of a single C$_{60}$ remains almost imperceptible. This result is in strong constrast with previous STM measurements in which a very important increase of LDOS was observed but only for the conduction band of a semiconducting peapod.\cite{yazdani} The slight variations in valence and conduction bands observed in the present study for semiconducting (and metallic) peapods suggest that an additionnal phenomenom could occur in the STM experiment and may explain the discrepancy. For example, a Coulomb blockade event or a structural deformation of carbon peapod induced by the STM tip can alter significantly the spectroscopic signature. There is clearly a possible blockade in that the charging energy of C$_{60}$ exceeds 270 meV, which is much larger than $kT$.\cite{cb} \\

The transport properties of the (A) metallic (10,10) and (B) semiconducting (16,0) nanotubes are not very much altered by the encapsulation of C$_{60}$'s. The main panels of Figure 4 shows the variation of the transmission function $\bar{T}(E)$ (or transmittance) for perfect and C$_{60}$ filled nanotubes as a function of electron energy. We reproduce also the DOS curves of CNTs and peapods (lower panels) in Fig. 4 to identify  the energy position where changes are observed. For the metallic tube, the presence of transmittance peaks, instead of a plateau, near $E_F$ is mainly due to our finite model for which the bands still have a molecular (discrete) character. The imperfect gold-CNT contact induces an extra contact resistance (of $\approx$ 8 k$\Omega$) to the minimal resistance of 6 k$\Omega$ for a metallic (ballistic) nanotube, and lowers the transmittance at $E_F$ from 2 to $\approx$  0.8.\cite{note1} However, since the CNTs and peapods have similar gold-tube geometry, we are expecting a similar contact resistance. The regions where small changes in transmittance are observed agree well with the features observed in DOS associated to C$_{60}$ (see Figure 4A). As observed for the electronic structure properties, the small diameter of the (16,0) tube facilitates a slight improvement of orbitals mixing between the CNT and the C$_{60}$'s. As a result, the transport properties of the (16,0) tube are more altered than for the (10,10) tube, especially in the valence band. However, these changes remain quite weak and suggest it would be experimentally very difficult to differentiate between a perfect nanotube and a peapod, at least on the basis on their electronic and electrical properties.\\

The existence of the highly endothermic C$_{60}$@(14,0) peapod is highly improbable because of the large activation energy needed for encapsulation (see Table I). This highly deformed peapod is however quite rich in information. The structure of  C$_{60}$@(14,0) contains a bumped CNT shell near C$_{60}$'s. Because of the small inner space in a (14,0) tube, the C$_{60}$ molecules are significantly squashed into an elliptic shape. Figure 5 shows the variation of the electronic (A) and electrical (B) properties between a perfect (14,0) tube and a 3C$_{60}$@(14,0) peapod. The DOS and LDOS diagrams (left panels) of the peapod show drastic variations with respect to the perfect (14,0) tube, more specifically in the band gap region where low energy bound states appear for the peapod. The influence of C$_{60}$ can now be clearly identified in the  LDOS curves over a large range of energy in both valence and conduction band. In addition, as the vertical dashed-dotted lines indicate, the variation of electronic structure of the (14,0) tube is mostly induced by structural deformations. This is deduced from a comparison of the perfect (14,0) tube and the empty peapod system. Considering the large charge transfer from C$_{60}$ to the tube (see Table 1 and Figure 3D), it is clear that C$_{60}$ has an important electronic and structural effect on the (14,0) tube properties in a peapod. This influence reflects on the transport properties where important fluctuations of the transmittance are observed near the first band edge of the conduction and valence bands. Nevertheless, the change of transmittance at $E_F$ is very weak and only a small difference is observed between the perfect tube and the peapod. Although we do not want to emphasize the electronic and transport properties of this endothermic peapod, the electronic structure of the squashed C$_{60}$ in this 3C$_{60}$@(14,0) peapod is useful to gain insight about the nature of the charge transfer involved in peapods.\\ 

In Figure 6, we compare the electronic structure of three different geometries of a C$_{60}$ triad, one with a ideal C$_{60}$ geometry (A), a second with the C$_{60}$ structure as in the (10,10) tube (B), and a third with the compressed C$_{60}$ structure as in the (14,0) tube (C). The symmetry group of the orbitals and their corresponding positions in the different arrangements are also included. The five-fold degenerate $h_u$ and three-fold degenerate $t_{1u}$ bands correspond to HOMO and LUMO, respectively.  Except for the small displacement of the HOMO and LUMO orbitals, and the small symmetry breaking of the $g_g + h_g$ manifold band at higher binding energy, the electronic structure of encapsulated-like C$_{60}$ molecules as in (10,10), is very similar to free C$_{60}$. On the other hand, the symmetry breaking of C$_{60}$ orbitals is very significant for the compressed geometry as in the C$_{60}$@(14,0) peapod. The more important band splitting is about the HOMO ($h_u$) where the symmetry breaking produces one four-fold degenerate band, and an isolated orbital that is strongly shifted toward $E_F$. This type of band splitting was already predicted for a single C$_{60}$ molecule by Joachim and coworkers.\cite{joachim} The presence of an occupied orbital near $E_F$ of the peapod increases significantly the ability of C$_{60}$ to donate electrons. In addition, as observed in LDOS of 3C$_{60}$@(14,0) peapod (see Figure 5), the low-lying energy states in the band gap of the (14,0) tube related to the regions of deformed CNTs, are placed in the appropriate range of energies to receive an extra electron from C$_{60}$. A charge transfer from the low-lying orbital of squashed C$_{60}$ near $E_F$ to the tube would then explain the important loss of charge density on the C$_{60}$ molecules observed near the nanotube wall reported in Table 1 and the large net positive charge on C$_{60}$ for the 3C$_{60}$@(14,0).\\

In summary, the electronic and electrical properties of metallic and semiconducting carbon nanotubes for the most stable carbon peapods are not significantly altered by the encapsulation of C$_{60}$. Only weak charge transfer are observed from the nanotube wall to the C$_{60}$ molecules. This change, mostly located between C$_{60}$, supports a weak orbital mixing between the two species. As the nanotube diameter decreases, within the exothermic peapods limit, a small increase in charge transfer and orbital mixing is observed. For the case of endothermic peapods, the changes in electronic and electrical properties are very drastic. The most important and relevant effect remains the C$_{60}$ symmetry breaking that induces the splitting of the HOMO ($h_u$ band) into several components especially one near $E_F$. This effect improves the electron donation ability of C$_{60}$ in the peapods.

\section*{acknowledgements}
I am grateful to RQCHP for providing computational resources. I am also pleased to acknowledge Richard Martel for his comments and helpful discussions, and Patrick Desjardins for his comments on the manuscript.

\newpage
\begin{center}
Table 1. Variation of the carbon peapods properties as a function of the nanotube diameter. \\
~~\\
\begin{tabular}{ccccc}
\hline \hline 
Peapod Type & ~~CNT Diameter~~ & ~d$_{\textrm{C}60-\textrm{C}60}$*~ & ~Mulliken Charge~ & $\Delta E_S$/C$_{60}$\\
 & (\AA) & (\AA) & (e/C$_{60}$) & (kcal/mol) \\
\hline  \hline
C$_{60}$@(15,15)   & 20.3 &   -     & -0.00 & -12\\
\hline
C$_{60}$@(10,10)   & 13.7 &   -     & -0.02 & -85\\
2C$_{60}$@(10,10) &         & 9.7 & -0.02 & -89\\
3C$_{60}$@(10,10) &         & 9.7 & -0.02 & -90\\
\hline  
C$_{60}$@(16,0)     & 12.5 &   -     & -0.07 & -78\\
2C$_{60}$@(16,0)   &         & 9.8 & -0.08 & -81\\
3C$_{60}$@(16,0)   &         & 9.8 & -0.08 & -83\\
\hline  
C$_{60}$@(9,9)      & 12.2  &   -     &-0.14 & -61\\
\hline
C$_{60}$@(15,0)     & 11.9  &   -     &-0.03 & -13\\
\hline
C$_{60}$@(14,0)     & 11.1 &    -    &+1.59 & 157\\
2C$_{60}$@(14,0)   &         &10.1 &+1.59 & 155 \\
3C$_{60}$@(14,0)   &         &10.1 &+1.59 & 155\\
\hline  
C$_{60}$@(8,8)      & 10.8  &   -     &+1.44 & 190\\
\hline \hline
\end{tabular}
\end{center}
\hskip1.5cm * distance between the center of adjacent C$_{60}$

\newpage
\begin{figure}[p]
\caption{Stabilization energy $\Delta E_S$ for the introduction of a C$_{60}$ molecule into a 100~{\AA} long carbon nanotube (H-terminated). Inner and outer diameter is when we considere the thickness of the $\pi$-electron cloud. Exothermic peapods have negative $\Delta E_S$, and endothermic peapods have positive values.}

\vskip0.5cm
\caption{Density of states (DOS) and local density of states (LDOS) of (A) metallic (10,10) and (B) semiconducting (16,0) nanotubes and peapods. DOS (upper panels) compares perfect (dotted line) and C$_{60}$ filled tubes (full line), while LDOS (lower panels) shows the contribution of a circular section of carbon atoms in the vicinity of a C$_{60}$ molecule (0 is directly over C$_{60}$, and $\pm$1, $\pm$ 2 is gradually away from C$_{60}$).}

\vskip0.5cm
\caption{Representation of the residual valence charge density of the 3C$_{60}$@$(n,m)$ peapods where (A) $n=m$=10, (B) $n=16, m=0$, (C) $n=m=9$, and (D) $n=14, m=0$. A negative value in the scale corresponds to a loss of charge density (bright region) while a positive indicates a gain of charge density (dark region). A similar scale is kept from (A) to (C) to emphasize the effect of the tube diameter on the charge density, and the scale for (D) is an order of magnitude higher.}

\vskip0.5cm
\caption{Variation of the electron transport (upper panels) and electronic (lower panels) properties from $(n,m)$ to 3C$_{60}$@$(n,m)$ systems, for (A) $n=m=10$ and (B) $n=16$, $m=0$. The electronic structure of a perfect $(n,m)$ tube (lower panels) is indicated by the dotted line-streaked area.}

\vskip0.5cm
\caption{Comparison of electronic structure (A) and electrical (B) properties between the $(14,0)$ and the 3C$_{60}$@(14,0) systems.}

\vskip0.5cm
\caption{Effect of space (tube) diameter on the electronic structure of (A) free C$_{60}$ when encapsulated in (B) (10,10), and (B) (14,0) CNT. Vertical lines show the origins and the displacement of splitted orbitals induce by symmetry breaking. }

\end{figure}

\newpage
\begin{figure}[p]
\vspace*{-1.0cm} 
\centerline{\includegraphics[angle=0,width=18cm]{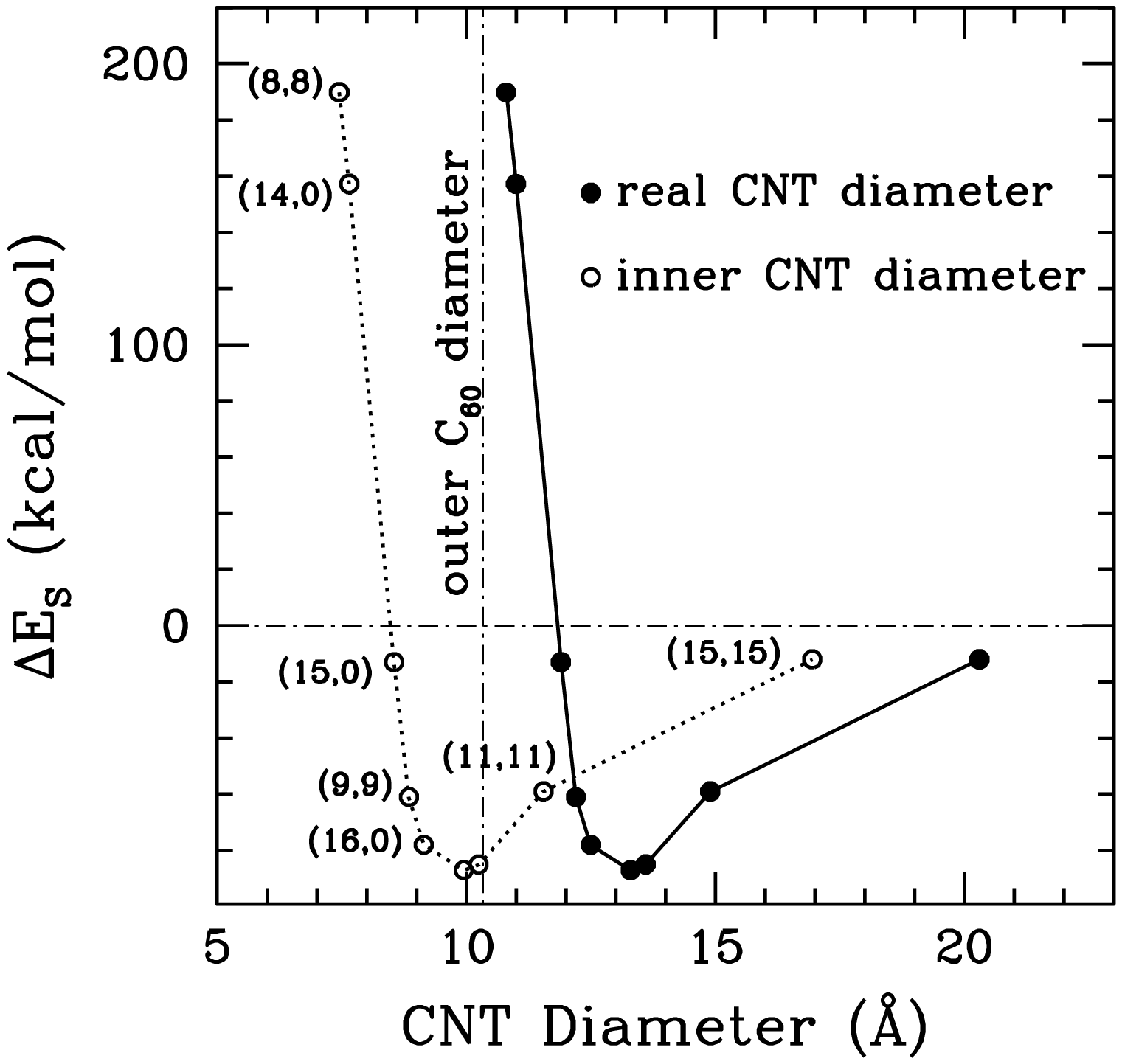}}
\vspace{1.0cm}
FIGURE 1. \textbf{Rochefort} \end{figure}

\newpage
\begin{figure}[p]
\vspace*{-4.0cm} 
\centerline{\includegraphics[angle=90,width=16cm]{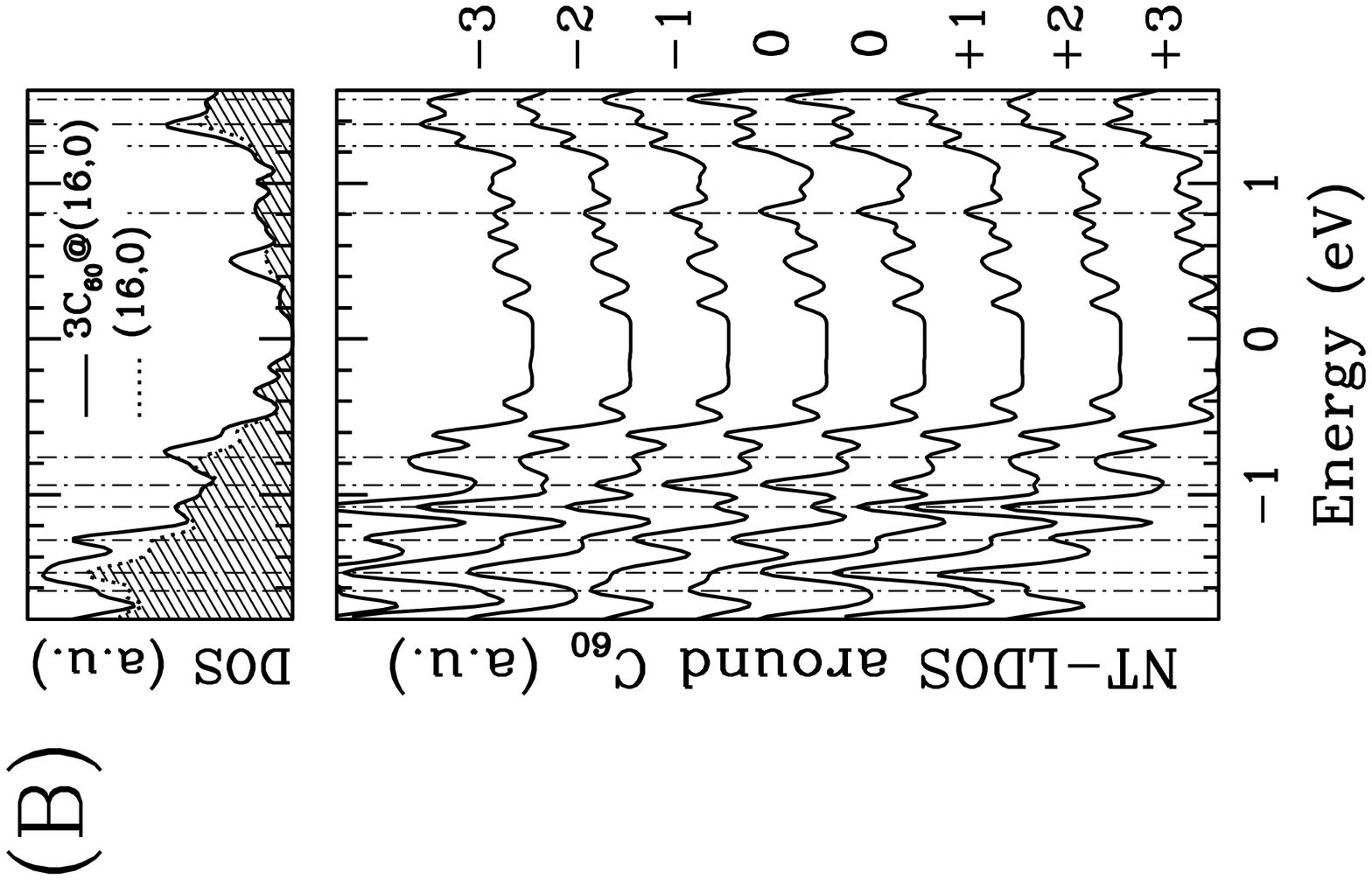}}
\vskip-6.0cm 
\centerline{\includegraphics[angle=90,width=16cm]{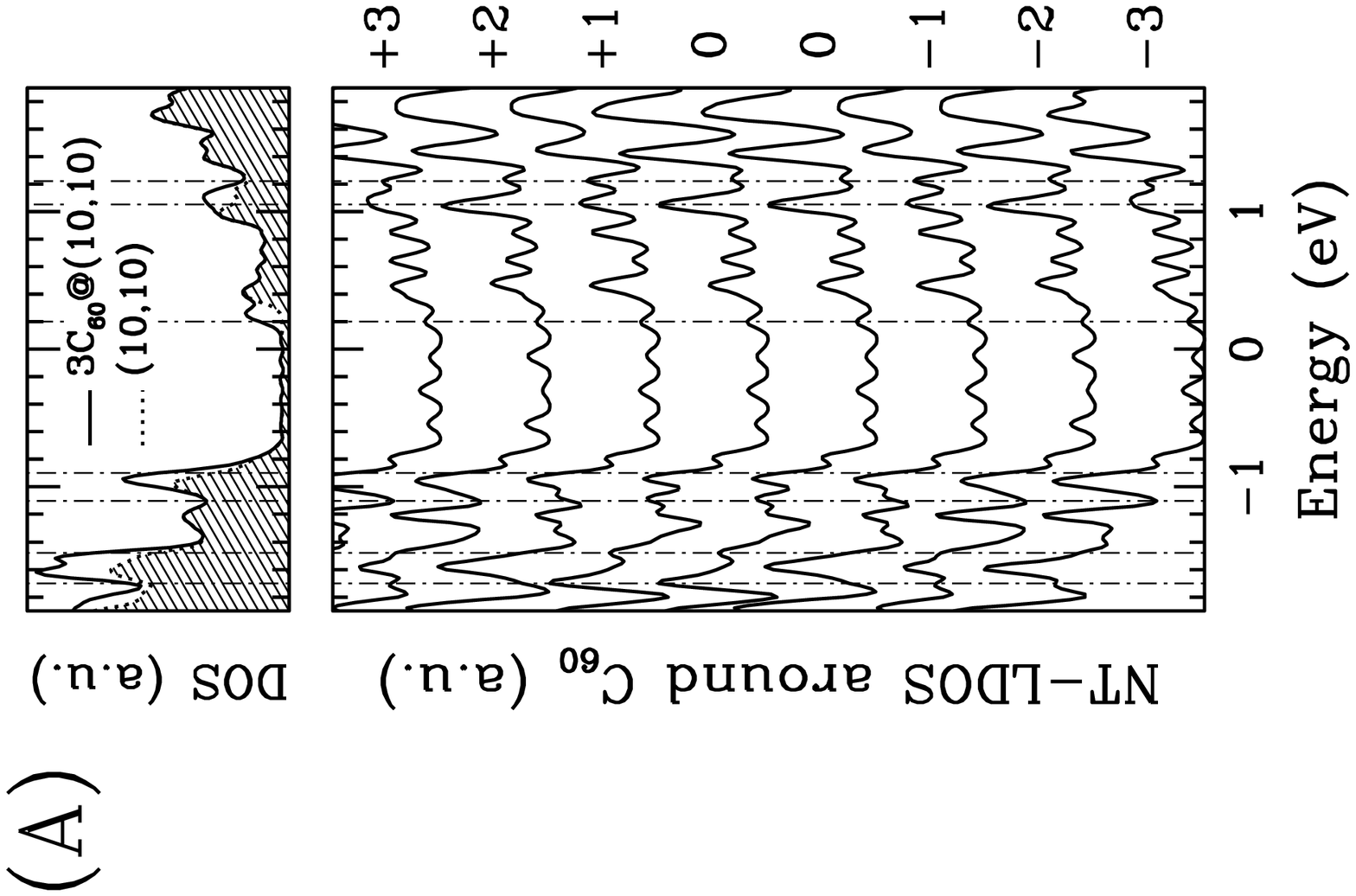}}
\vspace*{-2cm}
FIGURE 2. \textbf{Rochefort}
\end{figure}

\newpage
\begin{figure}[p]
\vspace*{1cm}
\centerline{\includegraphics[angle=0,width=8cm]{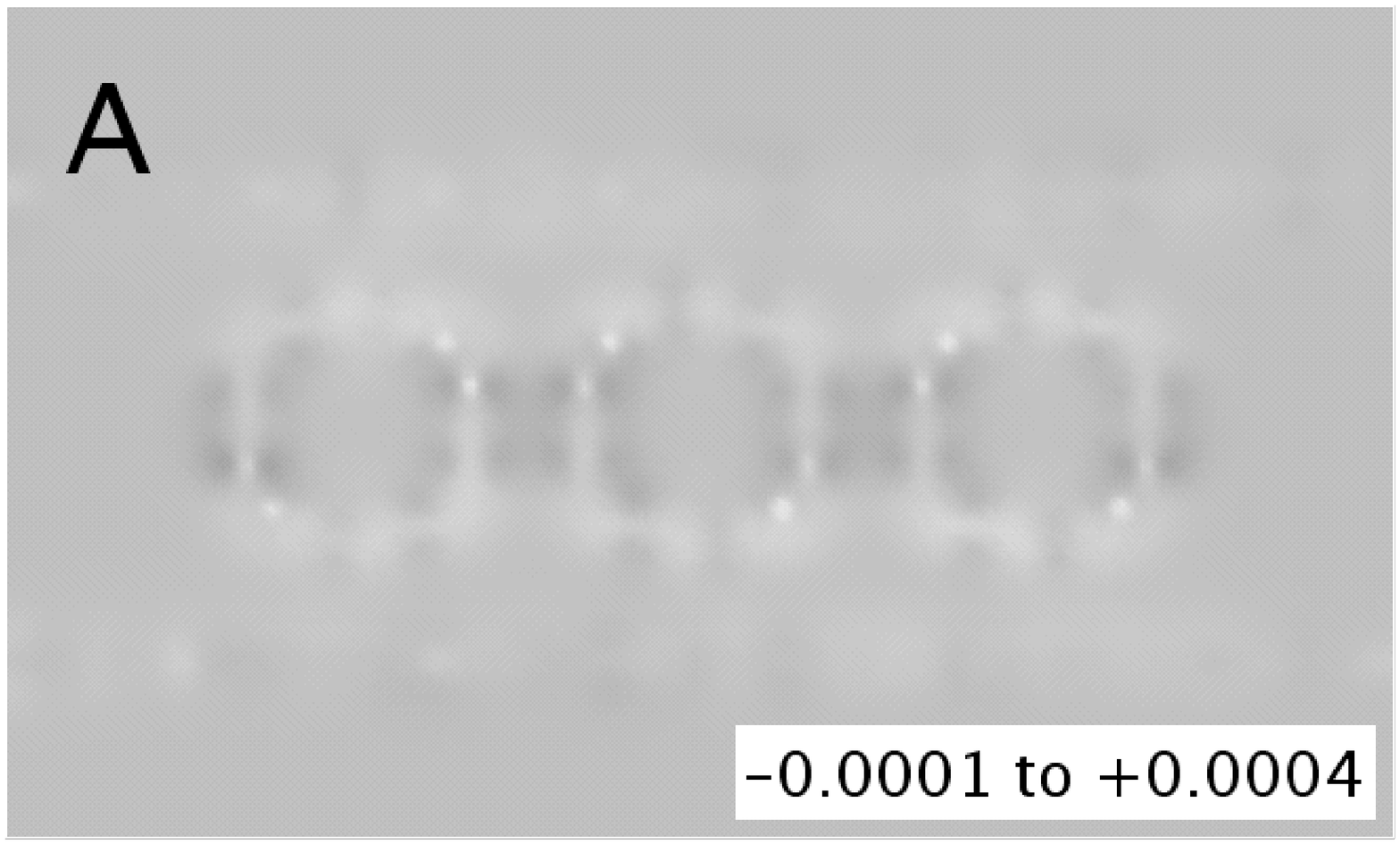}
\includegraphics[angle=0,width=8cm]{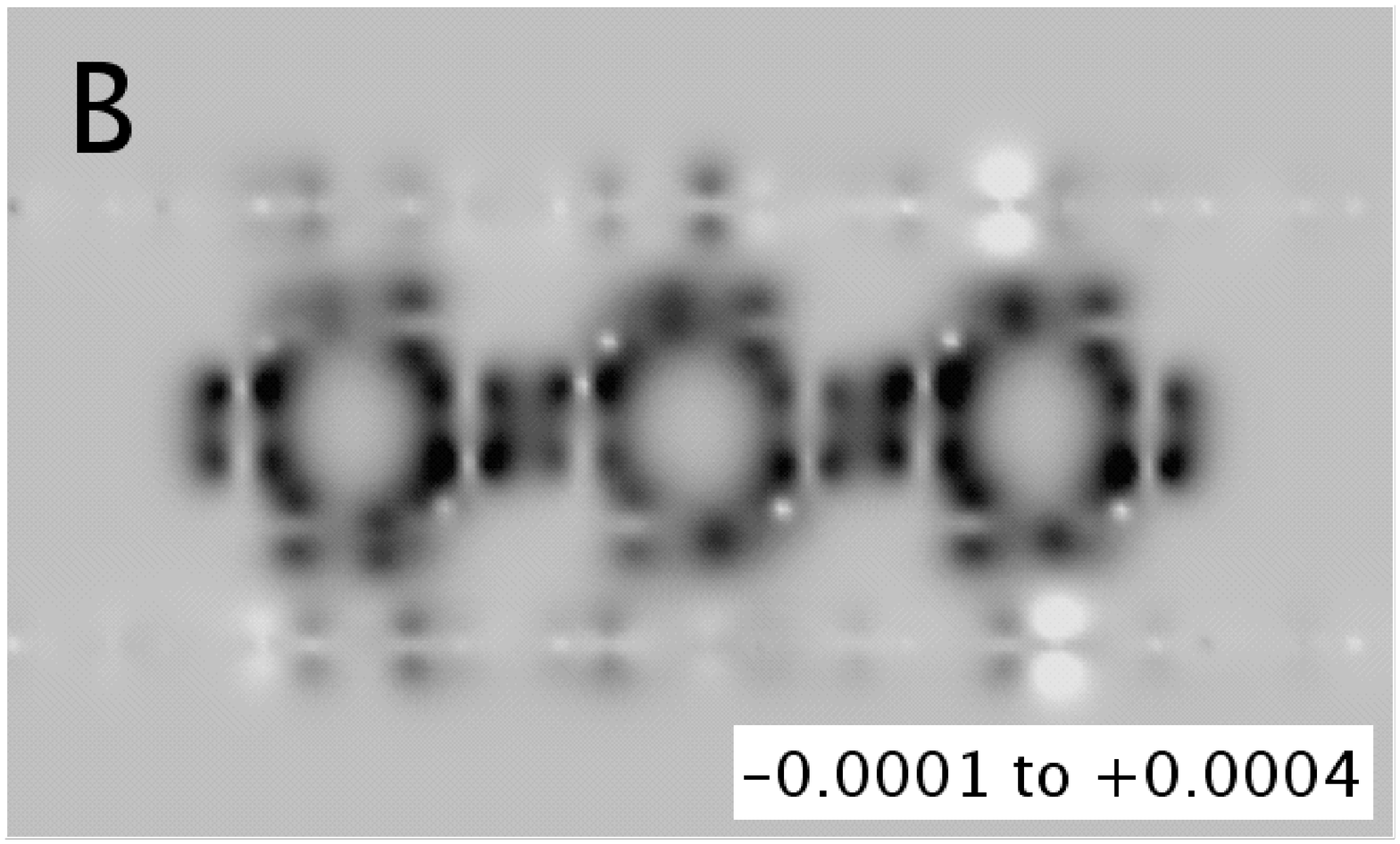}}
\vspace*{0.1cm}
\centerline{\includegraphics[angle=0,width=8cm]{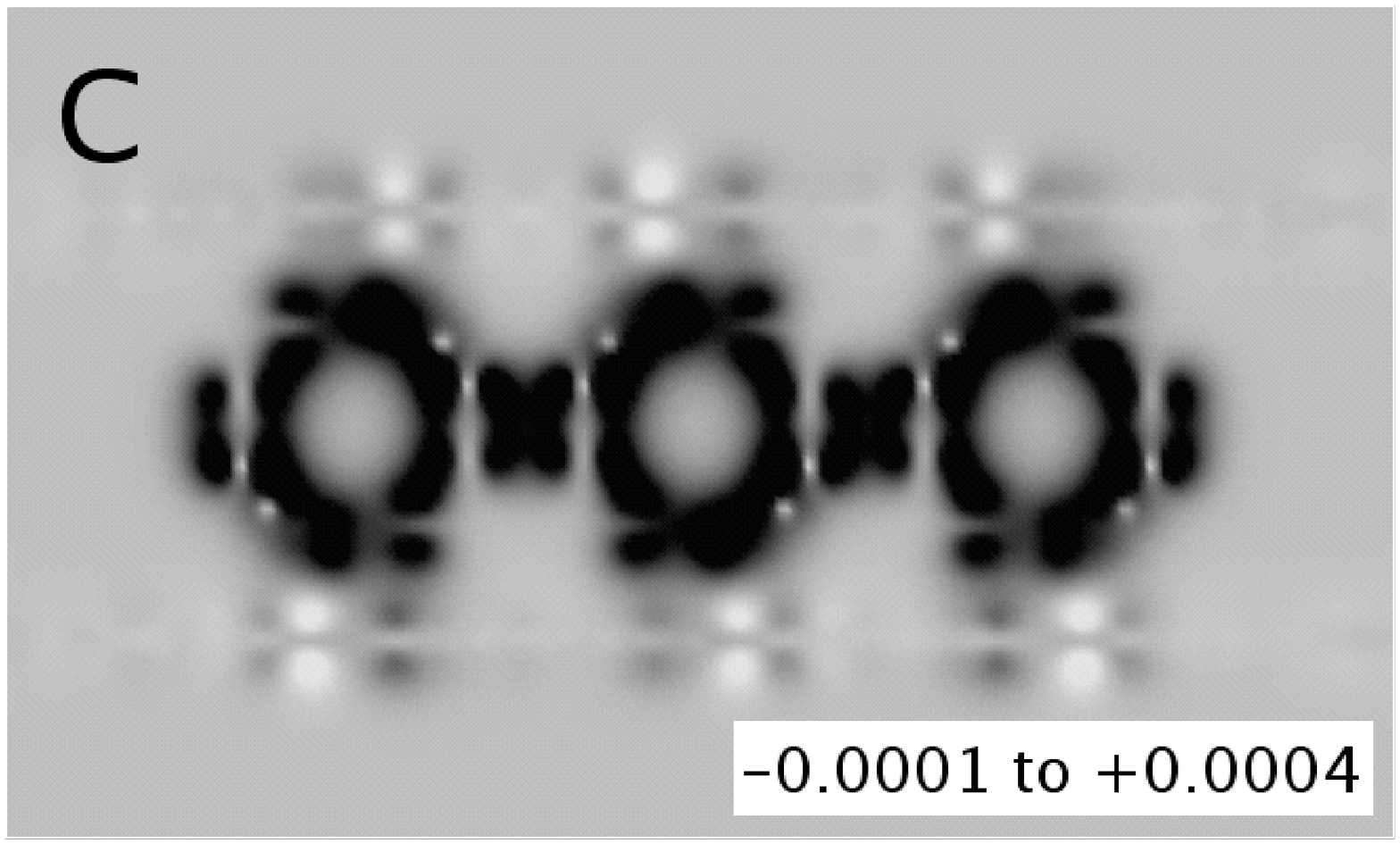}
\includegraphics[angle=0,width=8cm]{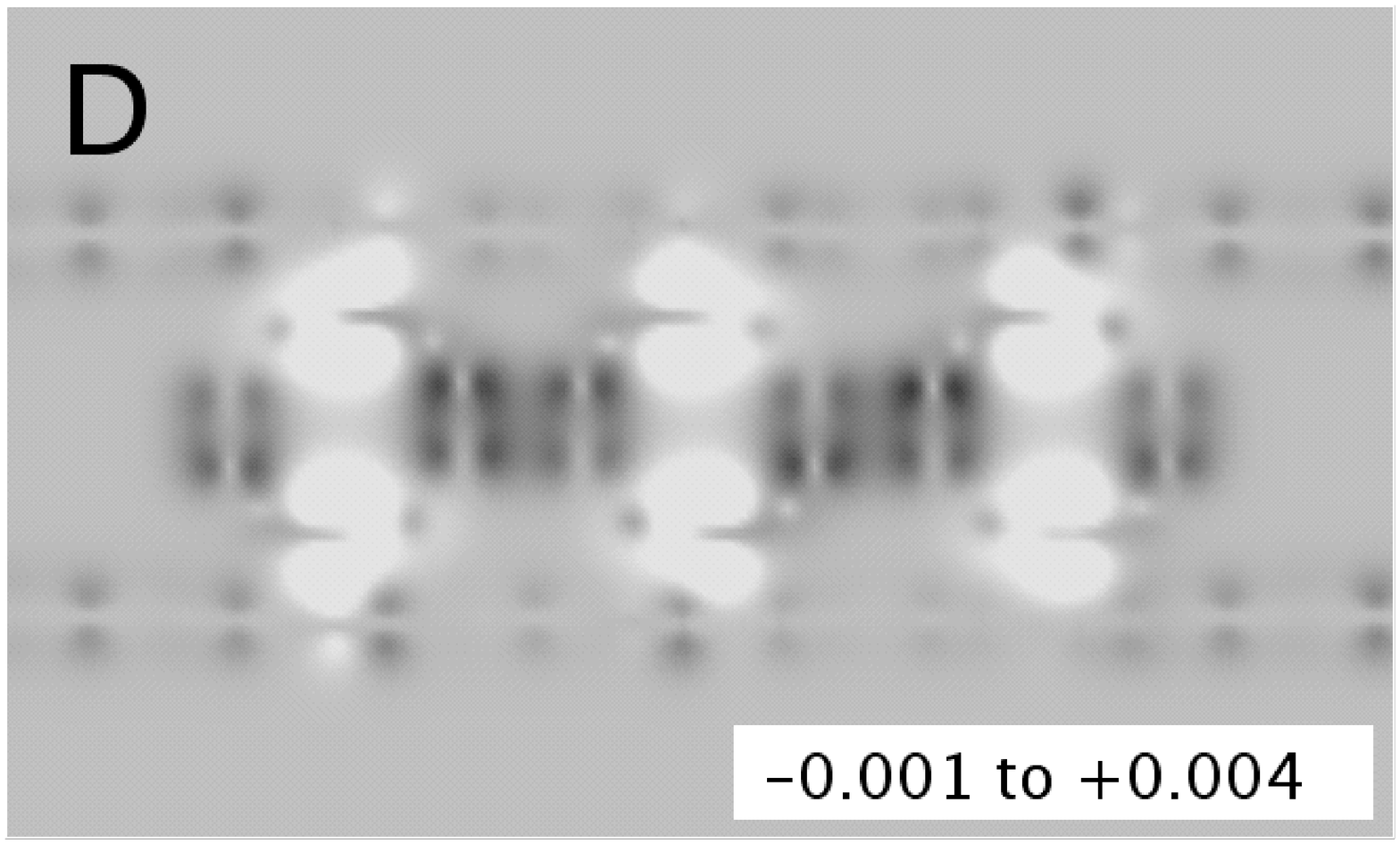}}
\vspace*{3cm}
FIGURE 3. \textbf{Rochefort} 
\end{figure}

\newpage
\begin{figure}[p]
\vspace*{-4cm}
\centerline{\includegraphics[angle=90,width=14cm]{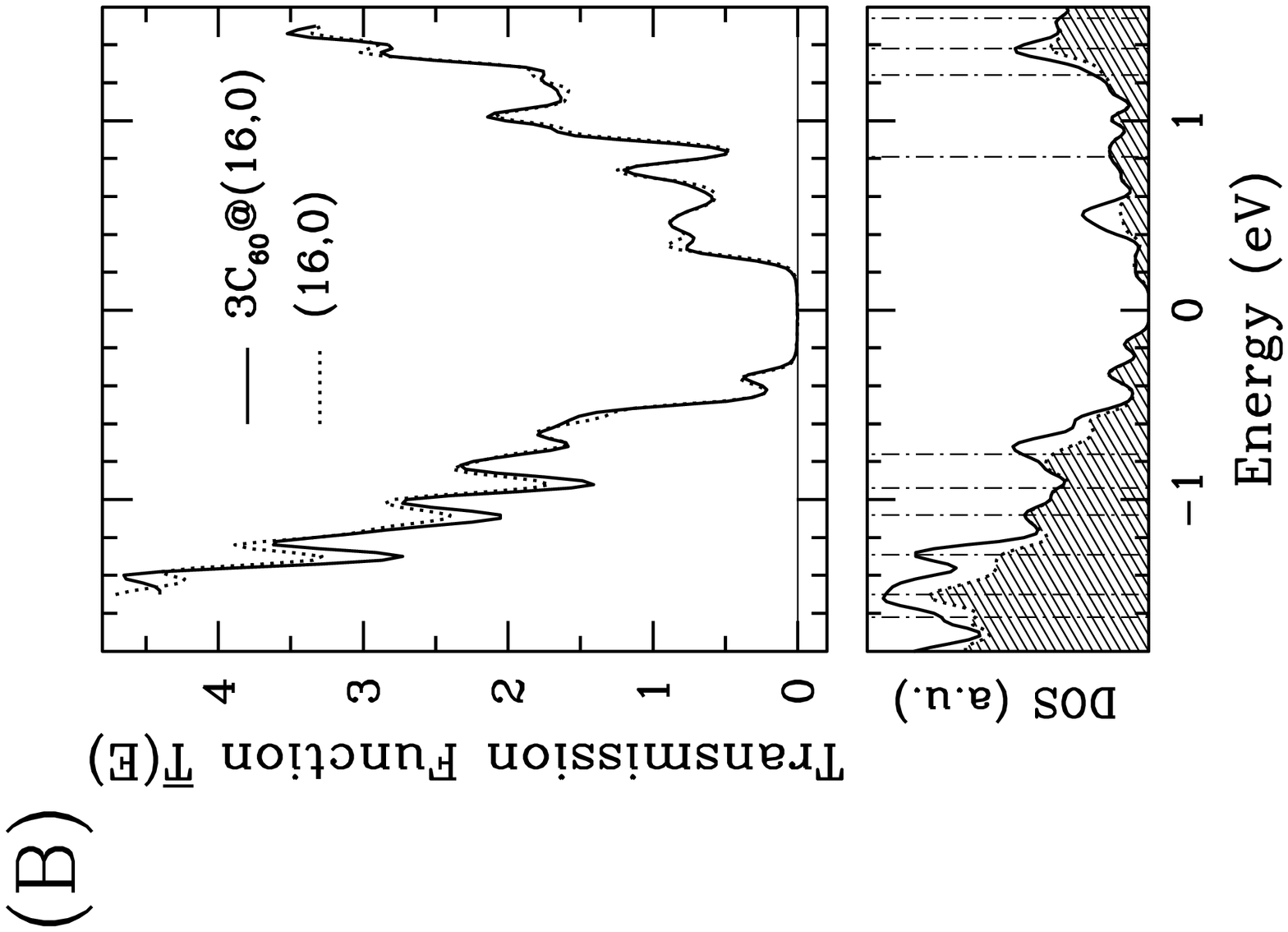}}
\vskip-3.5cm 
\centerline{\includegraphics[angle=90,width=14cm]{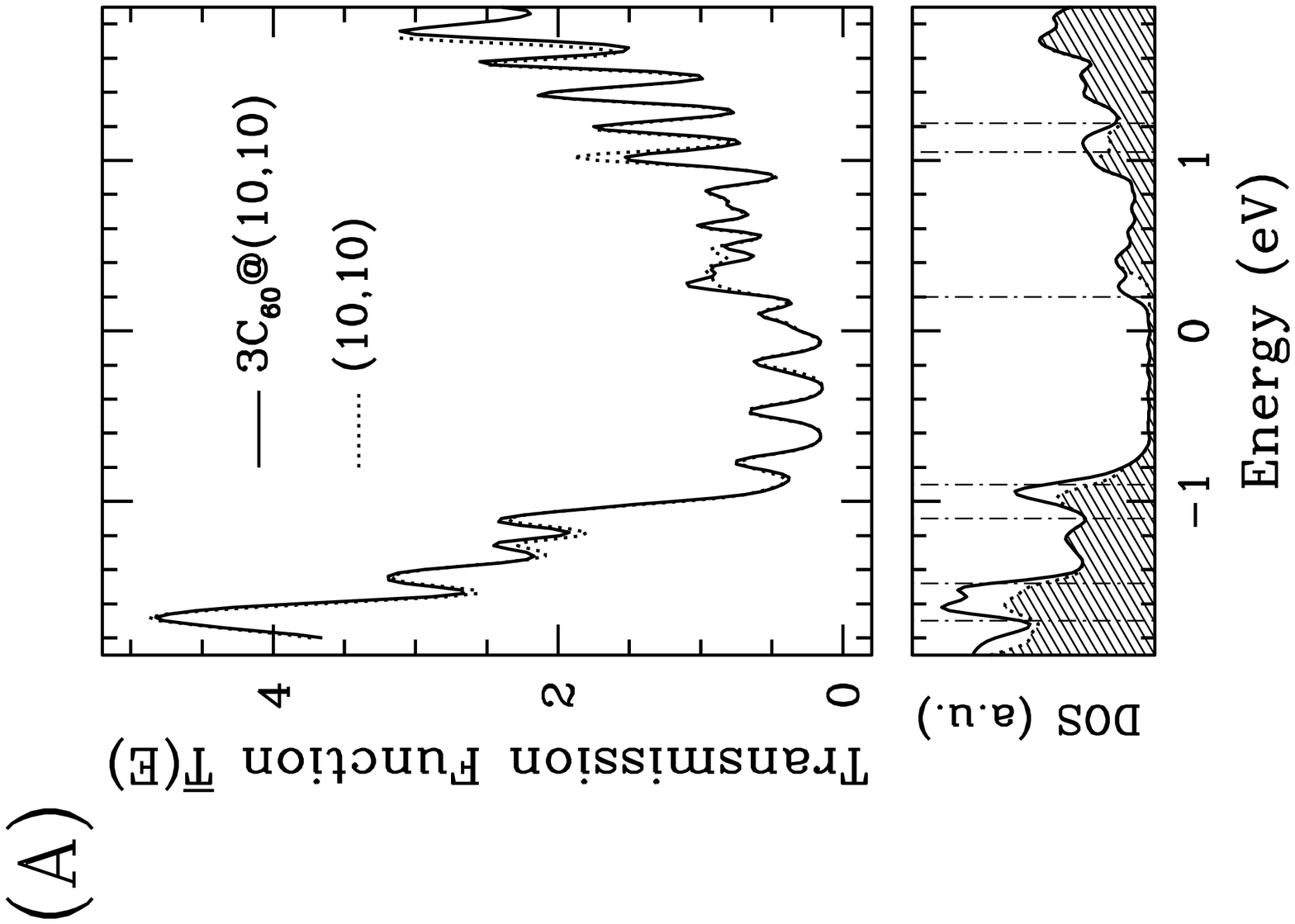}}
\vspace*{-1cm}
FIGURE 4. \textbf{Rochefort} 
\end{figure}

\newpage
\begin{figure}[p]
\vspace*{-3.5cm}
\hspace*{-1.0cm} \includegraphics[angle=90,width=16cm]{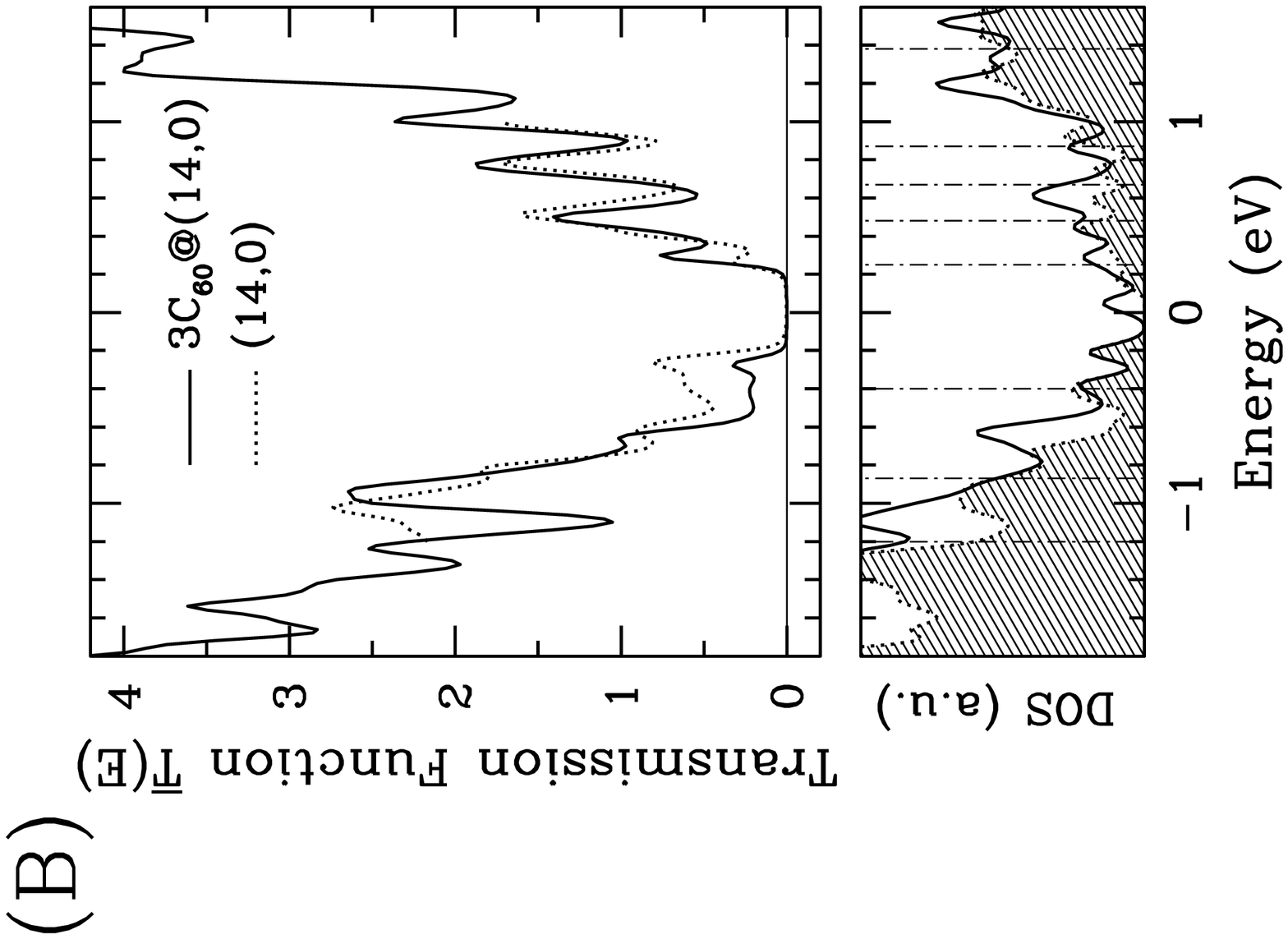}
\vskip-5.0cm 
\centerline{\includegraphics[angle=90,width=16cm]{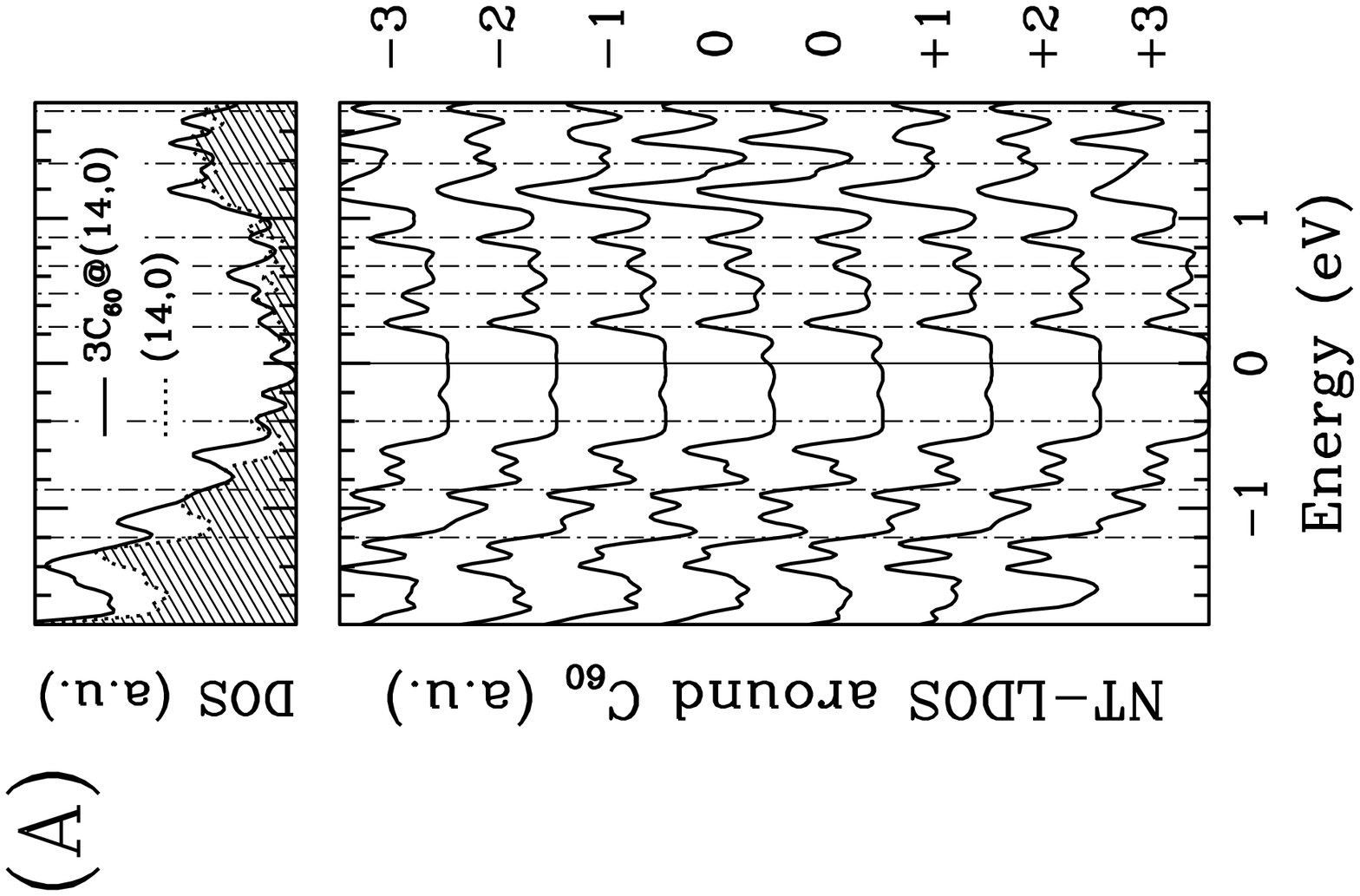}}
\vspace*{-2cm}
FIGURE 5. \textbf{Rochefort} 
\end{figure}

\newpage
\begin{figure}[p]
\vspace*{-1.0cm} 
\centerline{\includegraphics[angle=0,width=18cm]{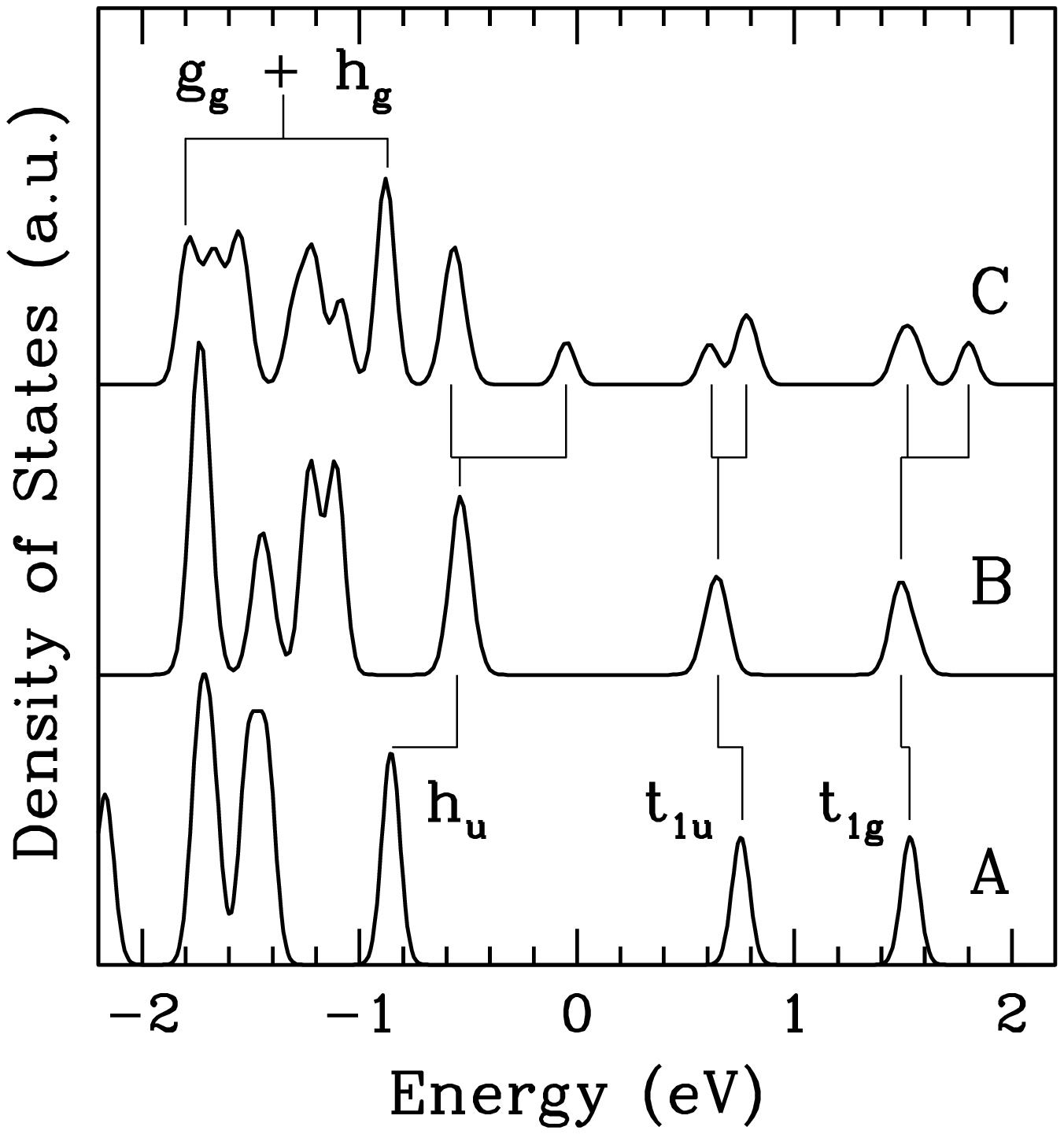}}
\vspace{1.0cm}
FIGURE 6. \textbf{Rochefort}
\end{figure}


\begin{references}
\bibitem{note} The net charge was evaluated from a Mulliken analysis of the
extended H\"uckel (EH) calculations. We are not pretending to determine a
quantitative charge distribution within EH theory, even if the present
description of a charge transfer is qualitatively correct.
\bibitem{note1} the extra-resistance of gold-nanotube contacts depends on the structural compatibility between the Au(111) crystal and the carbon atoms in the circular section which is the diameter. An extra-resistance of 5 k$\Omega$ was found for a (6,6) tube in \cite{rochefort_prb}
\end{references}
\end{document}